\documentclass[twocolumn,showpacs,preprintnumbers,amsmath,
amssymb,superscriptaddress,floatfix]{revtex4}
\usepackage{graphicx}% Include figure files
\usepackage{bm}% bold math

\def\mib#1{\hbox{\boldmath $#1$}}

\def\eq#1{Eq.\,(\ref{#1})}
\def\bfsigma{\mib{\sigma}}

\def\bk{\mib{k}}
\def\bn{\mib{n}}
\def\bp{\mib{p}}
\def\bq{\mib{q}}

\def\bK{\mib{K}}

\def\CG{{\cal G}}

\def\CV{{\cal V}}

\begin{document}

\preprint{APS/123-QED}

\title{$P$-wave $\Lambda N$-$\Sigma N$ coupling and the spin-orbit
splitting of $\hbox{}^9_\Lambda \hbox{Be}$}
% Force line breaks with \\

\author{Y. Fujiwara}
\email[]{yfujiwar@scphys.kyoto-u.ac.jp}
%\homepage[]{Your web page}
%\thanks{}
%\altaffiliation{}
\affiliation{
Department of Physics, Kyoto University,
Kyoto 606-8502, Japan
}
\author{M. Kohno}
%\email[]{}
%\homepage[]{Your web page}
%\thanks{}
%\altaffiliation{}
\affiliation{
Physics Division, Kyushu Dental College,
Kitakyushu 803-8580, Japan
}
\author{Y. Suzuki}
%\email[]{}
%\homepage[]{Your web page}
%\thanks{}
%\altaffiliation{}
\affiliation{
Department of Physics, and Graduate School
of Science and Technology,
Niigata University, Niigata 950-2181, Japan
}

\date{\today}% It is always \today, today,
             %  but any date may be explicitly specified

\begin{abstract}
We reexamine the spin-orbit splitting of $\hbox{}^9_\Lambda \hbox{Be}$ excited
states in terms of the $SU_6$ quark-model baryon-baryon interaction. 
The previous folding procedure to generate the $\Lambda \alpha$ spin-orbit
potential from the quark-model $\Lambda N$ $LS$ interaction kernel
predicted three to five times larger values
for $\Delta E_{\ell s}=E_x(3/2^+)-E_x(5/2^+)$ in the model FSS and fss2.
This time, we calculate $\Lambda \alpha$ $LS$ Born kernel,
starting from the $LS$ components of the nuclear-matter $G$-matrix
for the $\Lambda$ hyperon. 
This framework makes it possible to take full account of an important
$P$-wave $\Lambda N$--$\Sigma N$ coupling through the antisymmetric
$LS^{(-)}$ force involved in the Fermi-Breit interaction.
We find that the experimental value,
$\Delta E^{\rm exp}_{\ell s}=43\pm 5$ keV,
is reproduced by the quark-model $G$-matrix $LS$ interaction
with a Fermi-momentum around $k_F=1.0~\hbox{fm}^{-1}$,
when the model FSS is used in the energy-independent renormalized RGM
formalism.
On the other hand, the model fss2 gives too large
splitting of almost 200 keV. Based on these results and
the analysis of the Scheerbaum factors,
it is concluded that the model fss2 should be improved
to reproduce small single-particle
spin-orbit potentials of the $\Lambda$ hyperon.
\end{abstract}

% PACS, the Physics and Astronomy
                             % Classification Scheme.
.%Use showkeys class option if keyword
                              %display desired
\maketitle

%\end{document}

%\widetext

Study of hypernuclei based on the fundamental baryon-baryon
interactions is important, since the available scattering data
for the hyperon-nucleon ($YN$) interaction are very scarce.
The $SU_6$ quark-model (QM) baryon-baryon interaction developed by
the Kyoto-Niigata group is a comprehensive model for all
the octet-baryons ($B_8$), which is formulated
in the $(3q)$-$(3q)$ resonating-group method (RGM)
using the spin-flavor $SU_6$ QM wave functions,
a colored version of the one-gluon exchange Fermi-Breit interaction,
and effective meson-exchange potentials acting
between quarks \cite{PPNP}.
The early version, the model FSS \cite{FSS}
includes only the scalar (S) and pseudoscalar (PS) meson exchange
potentials as the effective meson-exchange potentials,
while the renovated one fss2 \cite{fss2,B8B8} introduces
also the vector (V) meson exchange potentials
and the momentum-dependent Bryan-Scott terms for the S and V mesons.
%Owing to these improvements, the model fss2 in the $NN$ sector
%has attained the accuracy comparable to that of one-boson exchange
%potentials.

As an important application of our QM baryon-baryon interactions,
we have carried out Faddeev calculations for the triton and
the hypertriton in Ref.\,\cite{ren}, in the most reliable framework
of using the energy-independent renormalized RGM kernels.
The triton binding energy predicted by fss2
is very close to the experimental value
with about 350 keV less bound, and the $\Lambda$ separation energy of the
hypertriton is 262 keV vs. the experimental value $130 \pm 50$ keV.
In the hypertriton calculation, the detailed information is
obtained for the central force of the $\Lambda N$ interaction,
since this system is $S$-wave dominant. On the other hand,
the information on the $\Lambda N$ $LS$ force is obtained,
for example, from the very small spin-orbit ($\ell s$) splitting
of the $5/2^+$ and $3/2^+$ excited states
of $\hbox{}^9_\Lambda \hbox{Be}$,
$\Delta E^{\rm exp}_{\ell s}=43 \pm 5~\hbox{keV}$ \cite{HT06}.
In the previous papers \cite{BE9L,2aljj}, we performed
Faddeev calculations of the two-alpha
plus $\Lambda$ ($\Lambda \alpha \alpha$) system by assuming
a simple $(0s)^4$ shell-model wave function for the $\alpha$ clusters.
For the $\alpha \alpha$ interaction, a microscopic $\alpha \alpha$ RGM kernel
is used with an effective $NN$ force, Minnesota three-range force.
The $\Lambda \alpha$ interaction is generated from
a simple two-range Gaussian central potential (SB potential),
which simulates the $S$-wave phase-shifts of the
$\Lambda N$ interaction of fss2
with a slight modification to fit the $\Lambda \alpha$ bound state.
The Pauli forbidden states between the two $\alpha$ clusters are
exactly eliminated in the three-cluster Faddeev formalism using
two-cluster RGM kernels \cite{TRGM,RED}.

The origin of the $5/2^+$ and $3/2^+$ splitting
in the $\Lambda \alpha \alpha$ cluster model
is the spin-orbit potential between $\Lambda$ and one of the $\alpha$ clusters,
which is known to be very small due to the strong cancellation between
the symmetric ($LS$) and antisymmetric ($LS^{(-)}$) $LS$ forces of
the $\Lambda N$ interaction. As a first step, we directly used
in Ref.\,\cite{2aljj} the QM $\Lambda N$ $LS$ RGM kernel
to generate the $\Lambda \alpha$ $LS$ potential
by simple $\alpha$-cluster folding.
In this procedure, the QM $\Lambda N$ $LS$ interaction of FSS or fss2
predicted 3 to 5 times larger values for $\Delta E_{\ell s}$, which is
not much improved in comparison with the results of Nijmegen simulated
potentials \cite{Hi00}.
It was pointed out in Ref.\,\cite{2aljj} that a further reduction
is possible in the model FSS, if one can properly take into account
the short-range correlation of the $P$-wave $\Lambda N$--$\Sigma N$ coupling
by the $LS^{(-)}$ force. This was conjectured through the analysis
of the Scheerbaum factors for the single-particle (s.p.)~spin-orbit
potentials, calculated in the $G$-matrix formalism. 

In this new calculation, we generate $\Lambda \alpha$ $LS$ Born kernel
from the $LS$ component of the nuclear-matter $G$-matrix
for the $\Lambda$ hyperon.
For the $(0s)^4$ $\alpha$-cluster folding, a new method developed
in Ref.\,\cite{B8a} is employed to derive direct and knock-on 
terms of the interaction Born kernel from the hyperon-nucleon $G$-matrices
with explicit treatments of the nonlocality and
the center-of-mass (c.m.) motion between the hyperon and the $\alpha$ cluster.
The $G$-matrix calculations are carried out
by assuming a constant Fermi momentum $k_F$,
since the local density approximation 
does not seem to work in light nuclear systems.
The $G$-matrix equation is solved for the energy-independent
QM baryon-baryon interaction
formulated in the renormalized RGM \cite{3alpha, rRGM},
and the continuous prescription for intermediate spectra is employed.
A similar procedure of the energy-independent
renormalized RGM is also used for the $\alpha \alpha$ RGM kernel.

We start from the $\Lambda N$--$\Sigma N$ coupled-channel $G$-matrix
equation \cite{GMAT,SPLS}
\begin{eqnarray}
& & G_{\gamma \alpha}(\bp, \bq; K, \omega, k_F)
=V^{\rm RGM}_{\gamma \alpha}(\bp, \bq)
+\sum_\beta \frac{1}{(2\pi)^3} \int d\,\bk
\nonumber \\
& & \times V^{\rm RGM}_{\gamma \beta}(\bp, \bk)
\frac{Q_\beta(k, K)}{e_\beta(k, K; \omega)}
~G_{\beta \alpha}(\bk, \bq; K, \omega, k_F)\ ,
\label{fm1}
\end{eqnarray}
where $Q_\beta(k, K)$ stands for the angle-averaged Pauli operator
and $e_\beta(k, K; \omega)=\omega-E_{b}(k_1)-E_N(k_2)$ is
the energy denominator.
Here, we use the notation $\beta$ etc.~to specify $\Lambda N$ or
$\Sigma N$ channel and $b$ for the corresponding $\Lambda$ or $\Sigma$.
Explicit expressions for $Q_\beta$ and $k_i$ are given in Ref.\,\cite{GMAT}.
The s.p.~energy $E_b(k)$ is defined by
\begin{equation}
E_b(k)=M_b+\frac{\hbar^2}{2M_b} k^2+U_b(k)\ \ ,
\label{fm3}
\end{equation}
with $U_b(k)$ and $M_b$ being the s.p.~potential
and the mass for the baryon $b$, respectively.
The starting energy $\omega$ is a sum
of the s.p.~energies of two interacting baryons:
\begin{eqnarray}
\ \hspace{-10mm} & & \omega = E_{a}(q_1)+E_N(q_2) = M_a + M_N
\nonumber \\
\ \hspace{-10mm} & & +\frac{\hbar^2}{2(M_{a}+M_N)} K^2
+\frac{\hbar^2}{2 \mu_\alpha} q^2
+U_a(q_1)+U_N(q_2)\ ,
\label{fm4}
\end{eqnarray}
where $\bK$ and $\bq$ are the total and relative momenta
corresponding to the initial s.p.~momenta $\bq_1$ and $\bq_2$.
The s.p.~potentials $U_\Lambda(q_1)$, $U_\Sigma(q_1)$ and $U_N(q_2)$
are determined self-consistently in the standard procedure
by assuming a constant $k_F$.
The determination of $q_1$ and $q_2$ is discussed below, in relation
to the folding formula of the $\Lambda \alpha$ Born kernel.

We employ the energy-independent Born kernel \cite{3alpha,rRGM}
for the $\Lambda N$--$\Sigma N$ coupling:
\begin{eqnarray}
V^{\rm RGM}(\bp, \bq)
=V_{\rm D}(\bp, \bq)+G(\bp, \bq)+W(\bp, \bq)\ ,
\label{fm5}
\end{eqnarray}
with
\begin{eqnarray}
W=\frac{1}{\sqrt{N}}\left(T_r+V_{\rm D}+G\right)\frac{1}{\sqrt{N}}
-\left(T_r+V_{\rm D}+G\right)\ .
\label{fm6}
\end{eqnarray}
Here, $T_r$ is the kinetic-energy operator for the relative motion
and $N$ stands for the normalization kernel
in the $\Lambda N$--$\Sigma N$ space.
This energy-independent treatment of the
QM baryon-baryon interaction in the $G$-matrix
formalism requires some kind of orthogonalization procedure
for the $\Lambda N$--$\Sigma N$ $\hbox{}^1S_0$ channel,
since this channel involves a Pauli forbidden state
at the quark level. The redundant correction of the $G$-matrix
is carried out in a similar way to the RGM $T$-matrix used
in the Faddeev formalism.
The details will be published elsewhere \cite{gpot}.

The derived $G$-matrix interaction is expressed in the form of
the invariant $G$-matrices as \cite{B8a} 
\begin{eqnarray}
& & G_{\Lambda N, \Lambda N}(\bp, \bp^\prime; K, \omega, k_F)
=\langle \Lambda N\,|\,G(\bp, \bp^\prime;
K, \omega, k_F) \nonumber \\
& & \qquad \qquad - G(\bp, -\bp^\prime; K, \omega, k_F)\,P_\sigma\,P_\tau\,|
\,\Lambda N\,\rangle\ \ \nonumber \\
& & \qquad =g_0+g_{ss} (\bfsigma_1 \cdot \bfsigma_2)
+h_0\,i \widehat{\bn} \cdot (\bfsigma_1 + \bfsigma_2)
\nonumber \\
& & \ \qquad \qquad +h_-\,i \widehat{\bn} \cdot (\bfsigma_1 - \bfsigma_2)
+ \cdots\ .
\label{fm7}
\end{eqnarray} 
Here, $\widehat{\bn}=\bn/|\bn|$ with $\bn=[\bp^\prime \times \bp]$,
and the invariant functions, $g_0$ (central), $g_{ss}$ (spin-spin),
$h_0$ ($LS$), $h_-$ ($LS^{(-)}$), etc.
depend on $p=|\bp|$, $p^\prime=|\bp^\prime|$,
and $\cos \theta=(\widehat{\bp} \cdot \widehat{\bp}^\prime)$,
as well as the $G$-matrix parameters $K$, $\omega$ and $k_F$.
These are expressed by the partial-wave components
of the $G$-matrix as in Appendix D of Ref.~\cite{LSRGM}.
The spin-spin term and the omitted noncentral terms in \eq{fm7}
do not contribute to the $\Lambda \alpha$ Born kernel,
due to the spin-saturated property of the $\alpha$ cluster.

The $\Lambda \alpha$ Born kernel
\begin{eqnarray}
\ \hspace{-10mm}
V_{\Lambda \alpha}(\bq_f, \bq_i) & = & V^C(\bq_f, \bq_i)+V^{LS}(\bq_f, \bq_i)
~i \widehat{\bn}\cdot \bfsigma_\Lambda\ ,
\label{fm8}
\end{eqnarray}
is calculated by the folding formula derived in Ref.\,\cite{B8a}.
Here, $\widehat{\bn}=\bn/|\bn|$ with $\bn=[\bq_i \times \bq_f]$. 
For the angular momentum projection of the central and $LS$ terms,
it is convenient to use the momentum transfer $\bk=\bq_f-\bq_i$
and the local momentum $\bq=(\bq_f+\bq_i)/2$ of the $\Lambda \alpha$
system, together with the similar relationship
$\bk^\prime=\bp-\bp^\prime$ and $\bq^\prime=(\bp+\bp^\prime)/2$ at
the two-baryon level. For example, the central Born kernel
$V^C(\bq_f, \bq_i)$ in \eq{fm8} is expressed as
\begin{eqnarray}
\ \hspace{-10mm} 
& & V^C(\bq_f, \bq_i)= \CV^C(\bk, \bq)=4 e^{-\frac{3}{32\nu}k^2}
\left(\frac{2(1+\xi)^2}{3\pi \nu}\right)^{\frac{3}{2}}
\nonumber \\
\ \hspace{-10mm}
& & \times \int d \bq^\prime~\exp \left\{-\frac{2(1+\xi)^2}{3\nu}
\left(\bq^\prime-\frac{1+4\xi}{4(1+\xi)}\bq\right)^2\right\} \nonumber \\
\ \hspace{-10mm}
& & \times g_0 \left(\bq^\prime+\bk/2, \bq^\prime-\bk/2;
(1+\xi)|\bq-\bq^\prime|, \omega, k_F \right)\ ,
\label{fm9}
\end{eqnarray} 
where $\xi=M_N/M_\Lambda$
and $\nu$ is the harmonic oscillator size parameter
of the $\alpha$ cluster.
Transformations from $g_0$
%(\bp, \bp^\prime; K, \omega, k_F)$
to
$\CG_0(\bk^\prime, \bq^\prime)
=g_0(\bq^\prime+\bk^\prime/2, \bq^\prime-\bk^\prime/2;
(1+\xi)|\bq-\bq^\prime|, \omega, k_F)$
% $\left. k_F \right)$
and from $\CV^C(\bk, \bq)$ to $V^C(\bq_f, \bq_i)$ are
carried out numerically for their partial-wave components.
For the direct and knock-on terms like \eq{fm9},
we find $\bk=\bk^\prime$.
The relationship $K=(1+\xi)|\bq-\bq^\prime|$ in \eq{fm9} implies
that the local momentum $\bq$ of the $\Lambda \alpha$ Born kernel
corresponds to the initial momentum $\bq_1$ for the $G$-matrix
equation and $\bq^\prime$ to the relative momentum $\bq$
in \eq{fm4}.
The standard angular-averaging procedure for the $\bq^\prime$ integral
in \eq{fm9} gives the starting energy $\omega$ as a function
of $q_1$ and $q$, which we call $\omega(q_1, q)$ prescription.
We therefore assign $q_1 \rightarrow q=|\bq|$ and $q \rightarrow q^\prime
=|\bq^\prime|$ in \eq{fm9} and obtain a simple folding formula
for the partial-wave components \cite{B8a}.
In the application to the $n \alpha$ RGM using
the $G$-matrix $NN$ interaction \cite{nargm}, 
we have improved this method to make possible the treatment
of other interaction types.
This improved version specifies $\omega$ as a function of $q$ and $K$
(namely, $\omega=\omega(q, K)$) by applying the angular-averaging
procedure to $\bK$.  
The explicit angular-momentum projection on the $\bq^\prime$-dependence
in $K$ makes it possible to deal with the Pauli-forbidden state
in the $n \alpha$ relative motion.
In the following, we will also show the results
by this $\omega(q, K)$ prescription, but the difference
from the $\omega(q_1, q)$ prescription is only quantitative.

For the Faddeev calculation, we use the same conditions as
used in Ref.\,\cite{2aljj}, except for the exchange mixture
parameter $u$ of the SB $\Lambda N$ potential.
We here use $u=1$, which is the same value as in Ref.\,\cite{BE9L}.
The increase from $u=0.82$ to $u=1$ is
because the energy-independent treatment of the $\alpha \alpha$ RGM kernel
gives slightly more repulsive effect than the previous energy-dependent
treatment. With this $u$ value, the ground-state energy
of $\hbox{}^9_\Lambda \hbox{Be}$ is $-6.596 \sim -6.598$ MeV,
which corresponds to the experimental value $-6.62\pm 0.04$ MeV.
We have not used the central $\Lambda \alpha$ Born kernel
obtained from the $G$-matrix calculation, since the interaction
strength is rather sensitive to the assumed $k_F$ value.
For example, the $\Lambda \alpha$ bound-state energy, predicted
by FSS in the $\omega(q_1,q)$ (or $\omega(q, K)$) prescription,
is $-2.95$  ($-2.46$) MeV for $k_F=1.20~\hbox{fm}^{-1}$
and $-4.04$ ($-3.43$) MeV for $k_F=1.07~\hbox{fm}^{-1}$,
compared with the experimental value $-3.12 \pm 0.02$ MeV. 
The purpose of the present investigation is to examine
the $LS$ component from the QM $\Lambda N$--$\Sigma N$ interaction.

Table \ref{table1} shows the results of
Faddeev calculations in the $jj$-coupling scheme,
obtained by using the QM $G$-matrix $\Lambda \alpha$ $LS$ Born kernel.
The Fermi momenta $k_F=1.07,~1.20$, and $1.35~\hbox{fm}^{-1}$ correspond
to the densities $\rho=0.5\,\rho_0,~0.7\,\rho_0$, and $\rho_0$, respectively,
with $\rho_0=0.17~\hbox{fm}^{-3}$ being the normal saturation density.
The final values for the $\ell s$ splitting
of the $5/2^+$ and $3/2^+$ excited states
are $\Delta E_{\ell s}=39$ - 96 keV for FSS and  205 - 223 keV for fss2,
depending on the $k_F$ values in the range of 1.07 - 1.35 $\hbox{fm}^{-1}$.
If the $\omega(q, K)$ prescription is used, the results by fss2 are
similar, but those by FSS are 56 - 118 keV.
A smaller $k_F$ value gives a smaller $\ell s$ splitting.
If we compare these results with the experimental value
$\Delta E^{\rm exp}_{\ell s}=43\pm 5$ keV, we find that
the model FSS can reproduce the experimental value
if the $k_F$ value around 1.09 - 1.02 $\hbox{fm}^{-1}$ is used.
We find that the excitation energies of the $5/2^+$ and $3/2^+$ states
are almost 120 keV too low, when the $\ell s$ splitting is correctly
reproduced with FSS.
This is the result when the energy-independent 
renormalized RGM kernels are used for the $\alpha \alpha$ RGM kernel
and for the QM baryon-baryon interaction.
On the other hand, fss2 gives too large values around 200 keV.

\begin{table*}[htb]
\begin{center}
\caption{
The Scheerbaum factor $S_\Lambda$ for symmetric nuclear matter,
the Scheerbaum-like factor $\widetilde{S}_\Lambda$ from the
$\Lambda \alpha$ zero-momentum Wigner transform for the
spin-orbit force, and the energy splitting,
$\Delta E_{\ell s}=E_{\rm x}(3/2^+)-E_{\rm x}(5/2^+)$,
of the $\hbox{}^9_\Lambda \hbox{Be}$ excited states
predicted from the $\alpha \alpha \Lambda$ Faddeev calculations
using the QM $G$-matrix $\Lambda \alpha$ $LS$ Born kernel.
The model is fss2 and FSS and the continuous prescription is used
for intermediate spectra in the $G$-matrix calculation.
The energy-independent renormalized RGM kernel is used
for the $\alpha \alpha$ RGM kernel and for the QM
baryon-baryon interactions.
The $\omega(q_1, q)$ prescription is used for the starting energies.
The results by the $\omega(q, K)$ prescription are also shown
in the parentheses. (See the text.)
}
\label{table1}       % Give a unique label
% For LaTeX tables use
\renewcommand{\arraystretch}{1.2}
\setlength{\tabcolsep}{3mm}
\begin{tabular}{ccrrr}
%\hline\noalign{\smallskip}
\hline
 & $k_F~(\hbox{fm}^{-1})$ & 1.07 & 1.20 & 1.35 \\
 & $\rho/\rho_0$ & 0.5 & 0.7 & 1 \\
%\noalign{\smallskip}\hline\noalign{\smallskip}
\hline
$G$-matrix & fss2 & $-11.8$ & $-12.1$ & $-12.3$ \\
$S_\Lambda$ ($\hbox{MeV}\,\hbox{fm}^5$) & FSS
 & $-4.1$ & $-5.2$ & $-6.3$ \\
%\noalign{\smallskip}\hline\noalign{\smallskip} 
\hline
$\Lambda \alpha$ & fss2 & $-14.7$ ($-14.8$)& $-15.6$ ($-15.6$)
 & $-16.3$ ($-16.3$)\\
$\widetilde{S}_\Lambda$ ($\hbox{MeV}\,\hbox{fm}^5$) & FSS
 & $-3.5$ ($-4.6$)& $-5.7$ ($-6.7$) & $-7.6$ ($-8.7$) \\
\hline
$\Lambda \alpha \alpha$ Faddeev & fss2 & 205 (204) & 213 (214) 
 & 223 (220) \\
$\Delta E_{\ell s}$ (keV) & FSS 
 &  39 (56) &  68 (87) &  96 (118) \\
%\noalign{\smallskip}\hline\noalign{\smallskip}
\hline
\multicolumn{2}{c}{$\Delta E^{\rm exp}_{\ell s}$ (keV)}
& \multicolumn{3}{c}{$43 \pm 5$} \\
%\noalign{\smallskip}\hline
\hline
\end{tabular}
\end{center}
% Or use
%\vspace*{5cm}  % with the correct table height
\end{table*}

These results are consistent with the
tendency of the Scheerbaum factor $S_\Lambda$ in the nuclear matter.
Table \ref{table1} also lists the Scheerbaum factor $S_\Lambda$
in symmetric matter, indicating the strength
of the spin-orbit potentials of the $\Lambda$ hyperon.
A similar quantity can be derived for the zero-momentum
Wigner transform calculated from the $\Lambda \alpha$ $LS$ Born kernel
in \eq{fm8} (see Eq.\,(2.47) of Ref.\,\cite{B8a}).
This quantity, that we call the Scheerbaum-like
factor $\widetilde{S}_\Lambda$, is expected to give
a better measure for the strength of the $\Lambda \alpha$ spin-orbit
interaction, since it deals with the recoil
effect of the $\alpha$ cluster due to the correct treatment
of the c.m.~motion in the $\alpha$-cluster folding.
The recoil effect is about 20 - 30$\%$ and is by no means small,
as discussed in our previous paper \cite{2aljj}.
We find that the strong cancellation between the $LS$ and $LS^{(-)}$ forces
takes place in the QM Fermi-Breit interaction
for the $P$-wave $\Lambda N$--$\Sigma N$ coupling in the 
$\hbox{}^1P_1$--$\hbox{}^3P_1$ state, when the $G$-matrix equation is
solved especially in low-density nuclear matter.
This is most prominently exhibited in the model FSS. 
The spin-orbit contribution from the effective-meson exchange
potentials in fss2 does not lead to the small $\ell s$ splitting
of the $\Lambda$ hyperon, since the scalar-meson exchange $LS$ force
contains only the ordinary $LS$ and does not produce the $LS^{(-)}$ force.

The previous energy-dependent treatment of the RGM kernels yields
the results qualitatively similar to the present investigation.
The reduction of the energy splitting and the 
$S_\Lambda$, $\widetilde{S}_\Lambda$ factors for the smaller $k_F$
values is very drastic for FSS.
We find that $k_F=1.25~\hbox{fm}^{-1}$ will give the correct value
of $\Delta E_{\ell s}$ if FSS is used.
On the other hand, the model fss2 gives almost no reduction
for the smaller $k_F$ values.

In spite of the successful reproduction of the $\hbox{}^9_\Lambda
\hbox{Be}$ $\ell s$ splitting by the model FSS, there still remains an
important issue on the $P$-wave characteristics
of the $\Lambda N$ interaction.
Owing to the  very strong $P$-wave $\Lambda N$--$\Sigma N$ coupling in FSS,
the $\hbox{}^3P_1$ $\Sigma N$ resonance moves to the $\hbox{}^1P_1$
$\Lambda N$ channel, resulting in a very broad step-like resonance
in this channel, as seen in Fig.\,14 of Ref.\,\cite{PPNP}.
As the result, the cusp structure in the $\Lambda p$ total elastic
cross sections at the $\Sigma N$ threshold is largely enhanced
compared with that of the fss2 prediction, which is clearly
overestimated even from the present experimental data with large error
bars. See Fig.\,19 (e) of \cite{PPNP}.
The original $\hbox{}^3P_1$ $\Sigma N$ resonance is caused by the
attractive Pauli effect from the exchange kinetic-energy kernel,
related to the the Pauli forbidden $(11)_s$ $SU_3$ state for the 
most compact $(0s)^6$ six-quark configuration
in the flavor-symmetric channel.
The resonance behavior in the $\Lambda N$--$\Sigma N (I=1/2)$ 
$\hbox{}^1P_1$--$\hbox{}^3P_1$ state sensitively depends on
the strength of the $LS^{(-)}$ force and the strength of the
attractive central force in the $\Sigma N (I=1/2)$ channel.
Furthermore, the central $\Lambda N$ interaction of FSS has a problem
that the $\hbox{}^1S_0$ interaction is too attractive, in comparison
with the $\hbox{}^3S_1$ interaction. For this reason, the hypertriton
calculation in Ref.\,\cite{ren} leads to the large overbinding
when FSS is used. These inconsistencies between the central and
$LS$ components of the $\Lambda N$ interaction imply that we
still need better models to describe the $\Lambda$ hypernuclei
by means of the $SU_6$ QM baryon-baryon interaction.

Summarizing this work, we have carried out $\Lambda \alpha \alpha$
Faddeev calculations by employing the $\Lambda \alpha$ $LS$ Born kernel
generated from the $LS$ components of the nuclear-matter $G$-matrix
for the $\Lambda$ hyperon. One of our $SU_6$ QM baryon-baryon
interaction FSS can reproduce the very small $\ell s$ splitting
of $\hbox{}^9_\Lambda \hbox{Be}$ excited states,
$\Delta E^{\rm exp}_{\ell s}=43\pm 5$, when an appropriate $k_F$ value
corresponding to the half density of the normal saturation density
is employed in the $G$-matrix calculation.
The explicit value of $k_F$ depends on the model construction
even within the framework of the $\Lambda \alpha \alpha$ cluster model
for $\hbox{}^9_\Lambda \hbox{Be}$;
$k_F=1.09~\hbox{fm}^{-1}$ for the model FSS
with the $\omega(q_1, q)$ prescription and
$1.02~\hbox{fm}^{-1}$ with the $\omega(q, K)$ prescription,
when the energy-independent renormalized RGM kernels are used
for the $\alpha \alpha$ RGM kernel and for the QM baryon-baryon
interaction.
The previous energy-dependent version of the RGM kernel requires
$k_F=1.25~\hbox{fm}^{-1}$ to reproduce $\Delta E_{\ell s}$ by FSS.
On the other hand, the model fss2 gives too large splitting
of almost 200 keV.
An essential ingredient of the present formalism is to take into
account an important $P$-wave $\Lambda N$--$\Sigma N$ coupling
through the antisymmetric $LS^{(-)}$ force involved
in the Fermi-Breit interaction.
From the present results and the analysis of the Scheerbaum factors
for the s.p.~spin-orbit potentials, we conclude that the
spin-orbit contribution from the effective meson-exchange potentials
in fss2 needs to be improved to reproduce 
the small spin-orbit interaction of the $\Lambda$ hyperon,
experimentally observed.
Construction of a new model with consistent $\Lambda N$ central and
$LS$ interactions is now in progress.

\bigskip

\begin{acknowledgments}
This work was supported by Grants-in-Aids for Scientific
Research (C) (Grant Nos.~18540261 and 17540263),
and for Scientific Research on Priority
Areas (Grant No.~20028003),
and Bilateral Joint Research Projects (2006-2008)
from the Japan Society for the Promotion of Science (JSPS).
This work was also supported by the Grant-in-Aid
for the Global COE Program ``The Next Generation of Physics,
Spun from Universality and Emergence'' from the Ministry of Education,
Culture, Sports, Science and Technology (MEXT) of Japan. 
\end{acknowledgments}

\end{document}